\documentclass[pdflatex,sn-mathphys-num]{sn-jnl}

\usepackage{graphicx}%
\usepackage{multirow}%
\usepackage{amsmath,amssymb,amsfonts}%
\usepackage{amsthm}%
\usepackage{mathrsfs}%
\usepackage[title]{appendix}%
\usepackage{xcolor}%
\usepackage{textcomp}%
\usepackage{manyfoot}%
\usepackage{booktabs}%
\usepackage{algorithm}%
\usepackage{algorithmicx}%
\usepackage{algpseudocode}%
\usepackage{listings}%

\theoremstyle{thmstyleone}%
\theoremstyle{thmstyletwo}%

\theoremstyle{thmstylethree}%

\usepackage{graphicx}%
\usepackage{multirow}%
\usepackage{amsmath,amssymb,amsfonts}%
\usepackage{amsthm}%
\usepackage{mathrsfs}%
\usepackage[title]{appendix}%
\usepackage{xcolor}%
\usepackage{textcomp}%
\usepackage{manyfoot}%
\usepackage{booktabs}%
\usepackage{algorithm}%
\usepackage{algorithmicx}%
\usepackage{algpseudocode}%
\usepackage{listings}%

\usepackage{setspace}

\usepackage{isotope}
\usepackage{graphicx,tabularx,booktabs,colortbl}
\usepackage{pgfplotstable}
\usepackage{subcaption}
\usepackage{natbib}
\usepackage[load-configurations=abbreviations, separate-uncertainty=true, multi-part-units=single]{siunitx}
\usepackage[T1]{fontenc}

\newcommand{\ac}{\isotope[225]{Ac}}
\newcommand{\tl}{\isotope[209]{Tl}}

\DeclareSIUnit\year{y}

\begin{document}

\title[]{Can $^{225}$Ac be imaged on a long axial field-of-view PET/CT with standard clinical settings?}


\author*[1,2]{\fnm{Lorenzo} \sur{Mercolli}}\email{lorenzo.mercolli@insel.ch}

\affil[1]{\orgdiv{Department of Nuclear Medicine}, \orgname{Inselspital, Bern University Hospital, University of Bern}, \orgaddress{\city{Bern}, \postcode{CH-3018}, \country{Switzerland}}}

\affil[2]{\orgdiv{Albert Einstein Center for Fundamental Physics (AEC), Laboratory for High Energy Physics (LHEP)}, \orgname{University of Bern}, \orgaddress{\city{Bern}, \postcode{CH-3012}, \country{Switzerland}}}


\abstract{

\textbf{Purpose:} The $^{225}$Ac decay chain includes $^{209}$Tl, whose prominent 1567 keV gamma line can create electron–positron pairs. We investigated whether a long axial field-of-view (LAFOV) PET/CT can image $^{225}$Ac using standard clinical acquisition and reconstruction methods.

\textbf{Methods:} A syringe placed next to a vial (5.89 and 10.29 MBq of $^{225}$Ac activity) and a NEMA IEC body phantom (sphere activity concentration 4.56 kBq mL$^{-1}$) were scanned for 45 and 60 min, respectively, on a Biograph Vision Quadra (Siemens Healthineers). To support our findings, FLUKA simulations scored the annihilation locations around a $^{225}$Ac point source.

\textbf{Results:} The contours of the syringe and vial were visible despite high noise, but the NEMA spheres could not be distinguished from the background. As illustrated by the FLUKA simulations, the number of annihilation events decreases exponentially according to the mean free path of the 1567 keV photon in the surrounding medium. 

\textbf{Conclusion:} A LAFOV PET/CT can detect positrons created by pair production of the main $^{209}$Tl gamma line. However, these positrons are created where the photons interact with the surrounding medium rather than at the location of the $^{209}$Tl decays, which rules out standard PET imaging of $^{225}$Ac.
}


\maketitle
\newpage

\section{Introduction}

In recent years, the possibility to exploit the high linear energy transfer (LET) of $\alpha$ particles for cancer treatment has sparked interest in targeted $\alpha$ therapy (TAT)~\cite{Kim2012,JAMA2018,Jang2023,Tran2025,Somma2026}. Promising results from clinical trials have prompted significant efforts to translate $\alpha$-labeled tracers into clinical routine~\cite{Ninatti2025,Lim2025}. Apart from [\isotope[223]{Ra}]RaCl$_2$~\cite{Kluetz2014}, \ac\ is the $\alpha$ emitter that is closest to routine clinical use: \ac\ is produced at scale~\cite{Tosato2025}, early trials have been completed~\cite{Kratochwil2016,Emmett2026} and a phase II/III clinical trial (PSMAcTION, NCT06780670) is ongoing~\cite{Fendler2026}.

$\alpha$-emitting radionuclides for TAT often have a rather long decay chain, sometimes involving multiple $\alpha$ decays. It is therefore not surprising to find a rich structure of nuclear transitions with emission lines that can be used for single-photon emission computed tomography (SPECT) imaging. Indeed, with the exception of selective internal radiation therapy (SIRT) with \isotope[90]{Y}, therapeutic radionuclides are typically imaged with SPECT. However, SPECT imaging of $\alpha$ radionuclides for post-treatment verification and dosimetry is rather challenging~\cite{Seo2019,Wang2024}. Photon emission branching ratios are often small and the photon energies frequently lie outside the optimal range of a clinical SPECT/CT, being either too low (below \SI{100}{\kilo\electronvolt}) or too high (above \SI{400}{\kilo\electronvolt}).
Furthermore, quantitative SPECT imaging generally depends on the local dose calibrator, which for $\alpha$ emitters remains a challenge~\cite{alphamet}. The administered activity of radionuclides with multiple $\alpha$ decays in the decay chain is of $\mathcal{O} (\SI{10}{\mega\becquerel})$ or less, leading to very small activity concentrations in lesions and organs. 

SPECT imaging of \ac\ is feasible~\cite{Seo2019,Benabdallah2021,Wang2024,Tulik2024,Zhang2026}, since the two daughter nuclides \isotope[221]{Fr} and \isotope[213]{Bi} have prominent gamma lines at \SI{218.0\pm0.1}{\kilo\electronvolt} ($11.44\pm 0.12\%$ intensity) and at \SI{440.45 \pm0.01}{\kilo\electronvolt} ($25.9 \pm0.2\% $ intensity), respectively~\cite{Kondev2021}. 
The low activity concentration and septal penetration with medium-energy collimators (or, alternatively, the loss of sensitivity with high-energy collimators), however, make SPECT imaging of \ac\ far from straightforward. 

Compared with SPECT, positron emission tomography (PET) offers higher sensitivity, better spatial resolution and more robust absolute quantification. Post-therapy imaging is, however, typically not performed with PET due to the decay properties of therapeutic radionuclides. 
For SIRT, PET has proven valuable for post-treatment verification and dosimetry~\cite{Zeimpekis2023}, particularly with long axial field-of-view (LAFOV) PET/CT systems~\cite{Prenosil2022,Spencer2021,Zhang2024}. In the context of TAT, \isotope[149]{Tb} is an example of an $\alpha$-PET radionuclide~\cite{Umbricht2019,Mapanao2025,Mercolli2026b}, which, however, is still far from clinical translation. 

But would it be conceivable to image \ac\ with PET? After all, the decay chain of \ac\ contains \tl\ which emits a \SI{1567.08\pm0.02}{\kilo\electronvolt} line with a $99.663\pm0.007\%$ intensity. Being above \SI{1022}{\kilo\electronvolt}, these photons can produce electron-positron pairs which potentially lead to a signal in a PET detector. Ref.~\cite{Caravaca2026} recently showed that, by leveraging the time-of-flight (TOF) capabilities of modern PET/CT scanners, it is possible to image \ac\ with PET. However, the method of Ref.~\cite{Caravaca2026} requires the PET scanner to be operated in research mode with a modified energy range as well as a dedicated event selection and reconstruction method. In this study we investigate whether \ac\ can be imaged with PET without any special acquisition or analysis methods, simply by exploiting the high sensitivity of LAFOV PET/CT systems. At the same time, we show how the \tl\ emission properties and interaction properties of the \SI{1567.08}{\kilo\electronvolt} photon effectively preclude the use of PET for \ac\ without additional acquisition and analysis methods. 

\section{Methods}

We prepared two experimental setups: a syringe placed next to a vial and a standard NEMA IEC body phantom. The syringe was filled with \SI{5.89}{\mega\becquerel} of \ac\ in \SI{15}{\ml} and the vial contained \SI{10.29}{\mega\becquerel} in \SI{13.9}{\ml}. 
The activity concentration in the NEMA phantom spheres was \SI{4.56}{\kilo\becquerel\per\ml} while the background contained \SI{0.47}{\kilo\becquerel\per\ml}, i.e.\ the sphere-to-background activity concentration was about $9.7:1$.  

The syringe and vial were scanned for \SI{45}{\min} on a Biograph Vision Quadra (Siemens Healthineers, Knoxville, TN, USA) LAFOV PET/CT~\cite{Prenosil2022}, while the NEMA phantom was scanned for \SI{60}{\min}. The images were reconstructed with the vendor's ordered subset expectation maximization algorithm using the standard clinical settings, i.e.\ four iterations, five subsets, TOF, point-spread-function (PSF) recovery, attenuation correction, relative scatter correction and a \SI{2}{\mm} Gaussian post-reconstruction filter~\cite{Prenosil2022}. The matrix size was $440\times440$, resulting in a voxel size of \SI{1.65 \times 1.65 \times 1.65}{\mm\cubed}. The standard energy window \qtyrange{435}{585}{\kilo\electronvolt} was not modified. 

To interpret the images and to understand where positrons are generated, we performed two FLUKA~\cite{Ahdida2022,Battistoni2015,Donadon2024} simulations. First, we simply simulated an \ac\ point source in a very large water volume.  Second, we placed the point source in a spherical water volume (\SI{10}{\mm} radius) surrounded by air. A dedicated MGDRAW routine scored the spatial location of annihilation events. The radial distances from the point source were histogrammed.

\section{Results}

\begin{figure}
\centering
    \begin{subfigure}{0.32\linewidth}
        \includegraphics[width=\linewidth]{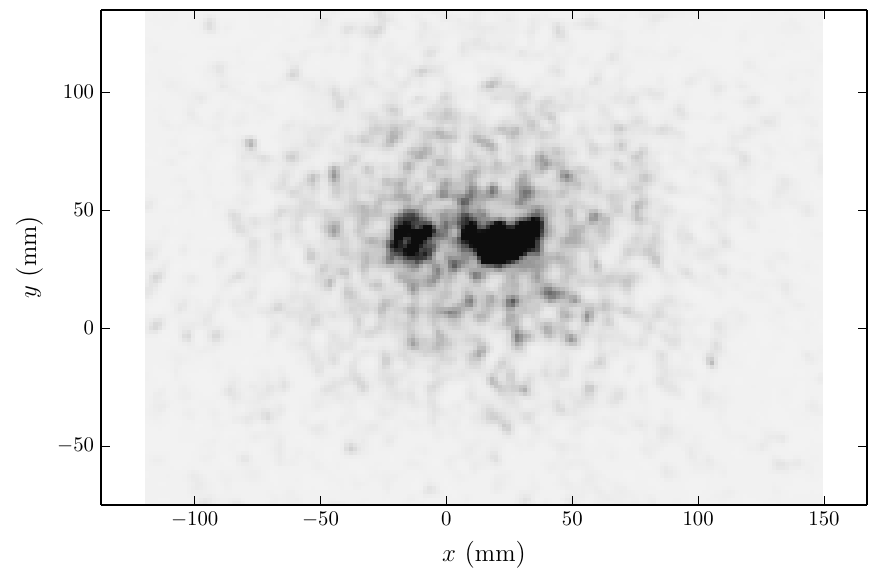}
        \caption{} \label{fig:slice_syringevial}
    \end{subfigure}
    \hfill
    \begin{subfigure}{0.32\linewidth}
        \includegraphics[width=\linewidth]{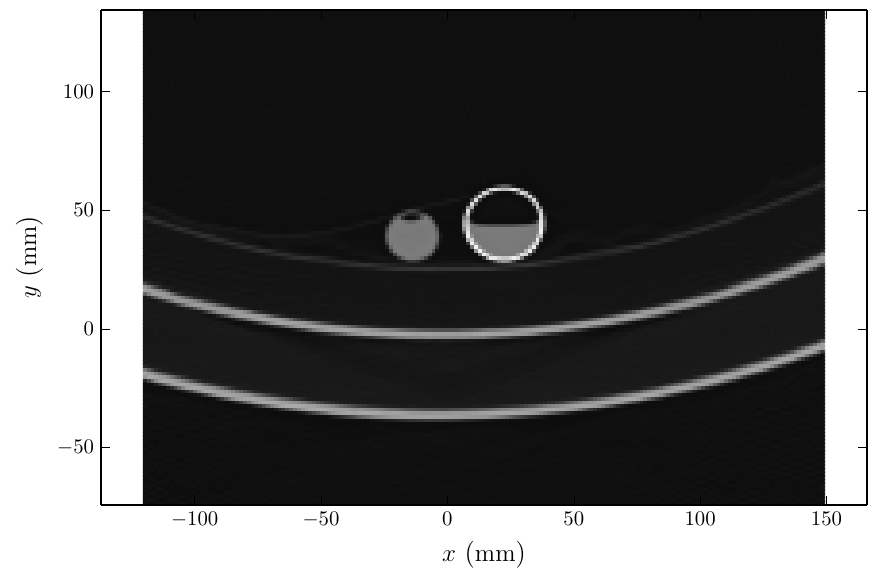}
        \caption{} \label{fig:slice_ct}
    \end{subfigure}
        \hfill
    \begin{subfigure}{0.32\linewidth}
        \includegraphics[width=\linewidth]{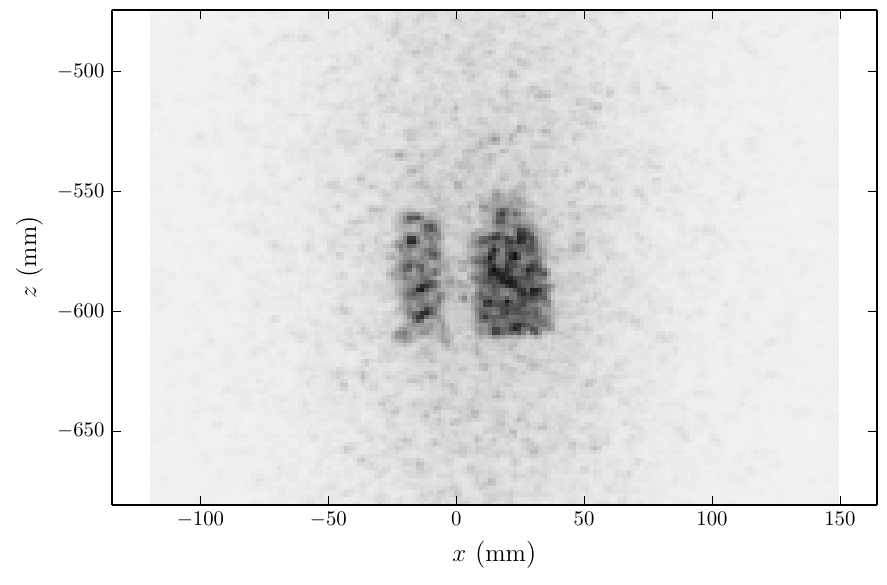}
        \caption{} \label{fig:mip_syringevial}
    \end{subfigure}
    \caption{Axial PET slice of the syringe next to the vial (\subref{fig:slice_syringevial}) and corresponding CT slice (\subref{fig:slice_ct}). Fig.~(\subref{fig:mip_syringevial}) shows the coronal MIP of the two objects. }  \label{fig:syringevial}
\end{figure}

\begin{figure}
\centering
    \begin{subfigure}{0.47\textwidth}
        \includegraphics[width=\linewidth]{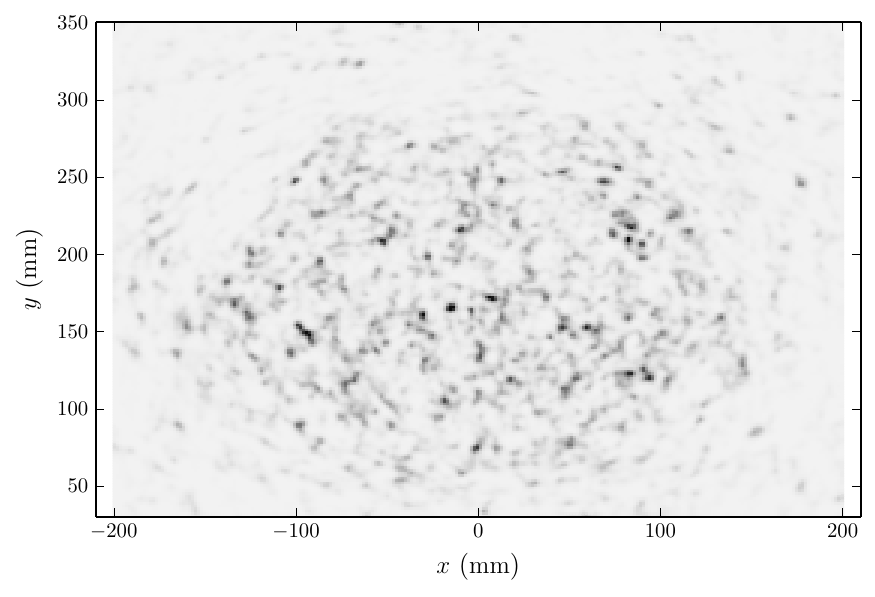}
        \caption{} \label{fig:slice_nema}
    \end{subfigure}
    \hfill
    \begin{subfigure}{0.47\textwidth}
        \includegraphics[width=\linewidth]{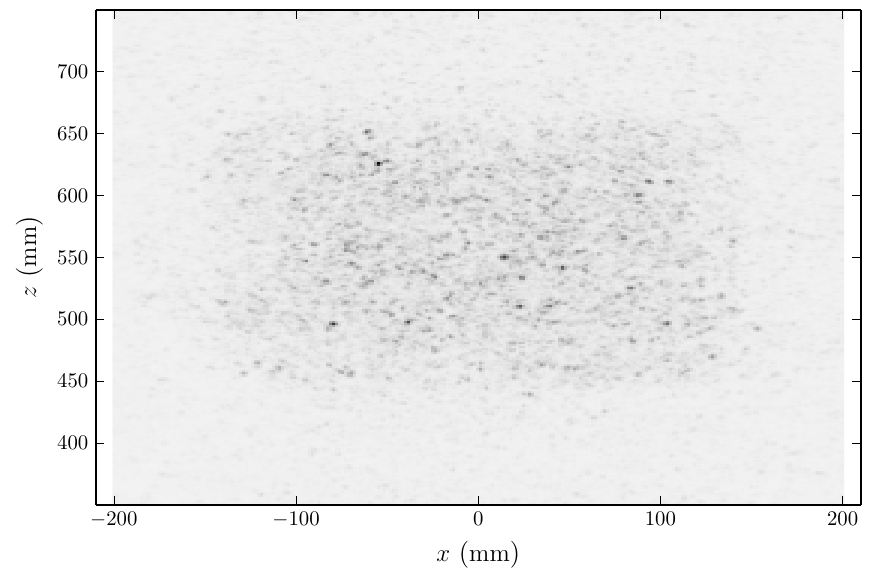}
        \caption{} \label{fig:mip_nema}
    \end{subfigure}
    \caption{Axial PET slice (\subref{fig:slice_nema}) and coronal MIP (\subref{fig:mip_nema}) of the NEMA phantom. }  \label{fig:nema}
\end{figure}

Fig.~\ref{fig:syringevial} shows an axial slice and the coronal maximum intensity projection (MIP) of the PET image of the syringe and vial next to each other. Although the image is very noisy, the contours of the syringe and vial are clearly visible. The corresponding images of the NEMA phantom are shown in Fig.~\ref{fig:nema}. The phantom's spheres are not discernible and the contours of the phantom are barely visible. The true coincidence count rate for the vial and syringe was about \SI{5}{\per\second} while for the NEMA phantom it was less than \SI{1}{\per\second}.

\begin{figure}
\centering
    \begin{subfigure}{0.49\textwidth}
        \includegraphics[width=\textwidth]{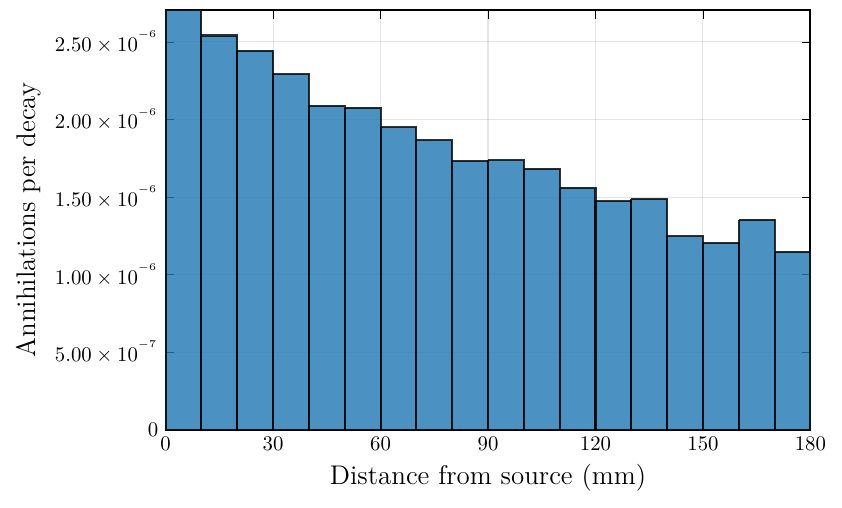}
        \caption{} \label{fig:hist_20cm}
    \end{subfigure}
    \hfill
    \begin{subfigure}{0.49\textwidth}
        \includegraphics[width=\textwidth]{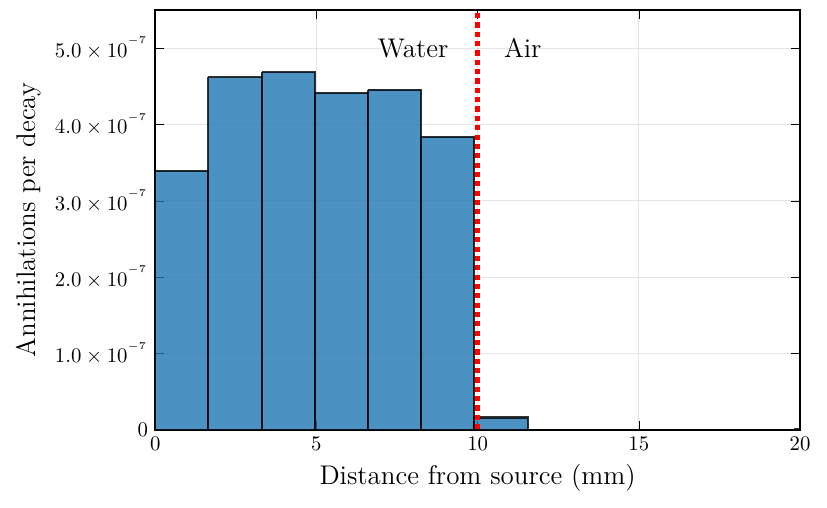}
        \caption{} \label{fig:hist_waterair}
    \end{subfigure}
    \caption{Histograms of the number of annihilations per \ac\ decay from FLUKA simulations. (\subref{fig:hist_20cm}) illustrates the exponentially decaying number of annihilation events. (\subref{fig:hist_waterair}) illustrates the sharp drop in annihilation events at the water-air boundary.}  \label{fig:histograms}
\end{figure}

Fig.~\ref{fig:histograms} illustrates the histograms of the annihilation locations from the FLUKA simulations. The radial bin width in Fig.~\ref{fig:hist_waterair} corresponds to the voxel size of the PET image.

\section{Discussion}

Fig.~\ref{fig:syringevial} demonstrates that Quadra can image annihilation photons from positrons created by pair production of (mainly) the \SI{1567.08\pm0.02}{\kilo\electronvolt} \tl\ gamma line. In equilibrium, the \tl\ activity is about $48$-times smaller than of \ac. Furthermore, the interactions of a \SI{1567.08}{\kilo\electronvolt} photon with water are dominated by Compton scattering. The pair-production cross section is about $585$ times smaller than that for Compton scattering~\cite{xcom}. From the FLUKA simulation in Fig.~\ref{fig:hist_waterair} we can already infer how small the positron production is (note that the volumes of the spherical bins in Fig.~\ref{fig:hist_waterair} are larger than the voxel volume in the PET images). 
Given these numbers, Fig.~\ref{fig:syringevial} is a remarkable showcase of Quadra's sensitivity, despite the image being noisy and the \SI{45}{\min} acquisition is much longer than a typical patient PET scan (though comparable to a \isotope[177]{Lu} post-therapy scan on a conventional SPECT/CT).

However, there are good reasons not to promote \ac\ imaging on LAFOV PET/CT systems with standard acquisition and reconstruction settings. First, Ref.~\cite{Caravaca2026} showed how to leverage the TOF capabilities of modern PET/CT systems to achieve accurate quantitative imaging of \ac\ even for low activity concentrations. In contrast to our study, however, the authors of Ref.~\cite{Caravaca2026} had to develop a dedicated prototype pipeline for both acquisition (adapting the detector gain to shift the energy window) and reconstruction. 
Fig.~\ref{fig:mip_nema} shows that, at low activity concentrations, Quadra reaches the limits of its sensitivity. Ref.~\cite{Caravaca2026} showed that the cascade $\gamma$-ray method can image activity concentrations comparable to those in the NEMA phantom spheres. However, sensitivity is actually not the main issue when imaging \ac\ with standard settings. 

Second, the fact that the positrons originate from pair production of the \tl\ gamma rays precludes quantitative imaging altogether. The \SI{1567.08}{\kilo\electronvolt} photons originate from the de-excitation of the $J^\pi = 5/2^+$ level of \isotope[209]{Pb}. This is an E2 transition and the internal pair production coefficient is $8.558\cdot 10^{-5}$~\cite{BrIcc}. This means that the creation of an on-shell photon of \SI{1567.08}{\kilo\electronvolt} dominates over internal pair production (unlike the case of the E0 transition and internal pair production of \isotope[90]{Zr} from the \isotope[90]{Y} decay).
However, the positrons from pair production of an on-shell photon are created at the location of the photon interactions in the medium through which it propagates. Hence, the location of the pair production follows the linear attenuation coefficient of the \SI{1567.08}{\kilo\electronvolt} photons. Fig.~\ref{fig:hist_20cm} shows how the number of annihilation events decreases exponentially with the distance from the source. The exponential decrease in Fig.~\ref{fig:hist_20cm} fits well the mean free path of \SI{1567.08}{\kilo\electronvolt} photons in water, which is \SI{17.71}{\cm}~\cite{NISTXray}. This means not only that in water the annihilation events are spread over large distances from the source, but also that the positron fraction depends on the surrounding material and not just on the physical properties of the decaying nucleus. In other words, the PSF in water is dominated by a long exponential tail with a full width at half maximum of about \SI{25}{\cm}. This clearly precludes both qualitative and quantitative PET imaging. Even if the activity concentration in the spheres were much higher, in the image they would be smeared into the phantom's background compartment.

That the contours of the vial and syringe can be identified in Fig.~\ref{fig:mip_syringevial} is due solely to the change in mean free path (attenuation coefficient) in the experimental setup. The mean free path of a \SI{1567.08}{\keV} photon in air is \SI{16.34}{\meter}~\cite{NISTXray}. Fig.~\ref{fig:hist_waterair} illustrates the sharp drop in number of annihilation events at the water-air interface (note that the vial's glass walls increase positron production, but this effect is not visible in the PET image). It is therefore clear that syringe and vial in Fig.~\ref{fig:mip_syringevial} are simply reflecting the change in the on-shell photon's mean free path rather than the spatial distribution of the \ac\ activity.

\section{Conclusions}

In this study, we showed that with standard clinical settings, a LAFOV PET/CT can detect annihilation events from the \SI{1567.08}{\keV} gamma line of \tl\ (one of the decay products of \ac). This illustrates the scanner's high sensitivity. However, internal pair production is negligible and therefore most positrons are created by pair production in the surrounding medium, far from the decaying nuclei. The resulting PET images reflect the attenuation properties of the surrounding material rather than the \ac\ distribution. Imaging \ac\ with PET requires dedicated methods, such as TOF cascade imaging demonstrated in Ref.~\cite{Caravaca2026}.

\backmatter

\section*{Acknowledgements}
We thank \^Angelo Felgosa Cardoso, Axel Rominger, Hasan Sari and Marco Viscione for useful discussions.

\section*{Declarations}

\bmhead{Funding}
This research is partially supported by the grant no. 216944 of the Swiss National Science Foundation.

\bmhead{Competing interests}
The author declares no competing interests.

\bmhead{Ethics approval}
Not applicable. 

\bmhead{Availability of data and materials}
The datasets generated and/or analyzed in the current study are available from the corresponding author upon reasonable request.

\bmhead{Consent to participate}
Not applicable.



\bibliography{ac225_bibliography}

\end{document}